\documentclass[a4paper,11pt]{article}
\pdfoutput=1 

\usepackage{jcappub} 
\usepackage{indentfirst}
\usepackage{appendix}
\usepackage{multirow}

\title{Cosmological parameter estimation with future gravitational wave standard siren observation from the Einstein Telescope}

\author[a]{Jing-Fei Zhang,}
\author[a]{Ming Zhang,}
\author[a]{Shang-Jie Jin,}
\author[a]{Jing-Zhao Qi,}
\author[a,b,c,1]{Xin Zhang\note{Corresponding author.}}

\affiliation[a]{Department of Physics, College of Sciences, Northeastern
University, Shenyang 110819, China}
\affiliation[b]{Ministry of Education Key Laboratory of Data Analytics and Optimization
for Smart Industry, Northeastern University, Shenyang 110819, China}
\affiliation[c]{Center for High Energy Physics, Peking University, Beijing 100080, China}

\emailAdd{jfzhang@mail.neu.edu.cn, 1800160@stu.neu.edu.cn, 850632821@qq.com, qijingzhao@mail.neu.edu.cn, zhangxin@mail.neu.edu.cn}

\abstract{In this work, we use the simulated gravitational wave (GW) standard siren data from the future observation of the Einstein Telescope (ET) to constrain various dark energy cosmological models, including the $\Lambda$CDM, $w$CDM, CPL, $\alpha$DE, GCG, and NGCG models. We also use the current mainstream cosmological electromagnetic observations, i.e., the cosmic microwave background anisotropies data, the baryon acoustic oscillations data, and the type Ia supernovae data, to constrain these models. We find that the GW standard siren data could tremendously improve the constraints on the cosmological parameters for all these dark energy models. For all the cases, the GW standard siren data can be used to break the parameter degeneracies generated by the current cosmological electromagnetic observational data. Therefore, it is expected that the future GW standard siren observation from the ET would play a crucial role in the cosmological parameter estimation in the future. The conclusion of this work is quite solid because it is based on the analysis for various dark energy models.}

\begin{document}
\maketitle
\flushbottom

\section{Introduction}
\label{sec1}

On 17 August 2017, the signal of gravitational waves (GWs) produced by the merger of a binary neutron star (BNS) was detected for the first time \cite{TheLIGOScientific:2017qsa}, meanwhile the signals of electromagnetic (EM) waves in various wave bands were also detected for the same transient source \cite{GBM:2017lvd}, which is fairly meaningful because the observations initiated the new era of multi-messenger astronomy.

In 1986, Schutz showed that the Hubble constant could be determined from GW observations \cite{Schutz:1986gp}. The observation of GWs from the merger of a binary compact-object system could give us the information of absolute luminosity distance, which could be considered as $standard$ $sirens$. It has the advantage of not using the cosmic distance ladder. Using only one GW standard siren data, we could determine the Hubble constant to be $H_0=70.0_{-8.0}^{+12.0}$ km s$^{-1}$ Mpc$^{-1}$ \cite{Abbott:2017xzu}, for which the error is still somewhat large. It is doubtless that future GW observations will provide precise constraints on the Hubble constant. In addition to this, the GW standard sirens could also be used to constrain various dark energy cosmological models.

Recently, some related issues about the future GW standard siren data have been discussed by some authors \cite{Cai:2016sby,Cai:2017aea,Cai:2017yww,Sathyaprakash:2009xt,Yang:2017bkv,Feeney:2018mkj,Zhao:2010sz,Liao:2017ioi,Wei:2018cov,Wei:2019fwp,Fu:2019oll,Yang:2019bpr,Yang:2019vni,Mendonca:2019yfo,Wang:2018lun,Zhang:2019ple,Zhang:2018byx,Zhang:2019ylr,Will:1994fb,Liu:2017xef,Berti:2018cxi,Zhao:2018gwk,Liu:2018sia,Li:2019ajo,Wang:2019tto}. For example, in Ref.~\cite{Zhang:2018byx}, it is shown that future GW standard siren observation could play a crucial role in breaking the parameter degeneracies generated by other observations and thus would significantly improve the cosmological parameter estimation in the future. However, it should be pointed out that there are two apparent drawbacks in the investigation of Ref.~\cite{Zhang:2018byx}: (i) That is only a preliminary investigation, because only two simplest dark energy models (the $\Lambda$CDM model and the $w$CDM model) are employed as examples in the analysis. (ii) In the simulation of the GW standard siren data, the central values of the cosmological parameters in the parameter planes are discrepant greatly from those from the conventional EM observations, which is disadvantageous in the analysis for how GW data improve the parameter estimation. Our work will overcome these drawbacks, and will make a more general analysis and give a more solid conclusion.


We consider more typical dark energy models in this work, including the $\Lambda$CDM model, the $w$CDM model, the Chevallier-Polarski-Linder (CPL) parametrization model \cite{MD:2001,EV:2003}, the $\alpha$ dark energy ($\alpha$DE) model \cite{c:2003}, the generalized Chaplygin gas (GCG) model \cite{MCB:2002}, and the new generalized Chaplygin gas (NGCG) model \cite{Zhang:2004gc}. We consider these models because they are typical and popular, and also according to the analysis in Ref.~\cite{Xu:2016grp} they are still relatively favored by the current observations. It should be noted that the holographic dark energy model has been discussed in the similar way in Ref.~\cite{Zhang:2019ple}, and thus we do not include this model in this work. In addition, some typical interacting dark energy models have also been discussed in Ref.~\cite{Li:2019ajo}.

To make an analysis for comparing the GW data and EM data on constraining cosmological parameters, we employ the mainstream cosmological probes based on EM observations, including the cosmic microwave background (CMB) data from the Planck 2018 mission \cite{Aghanim:2018eyx,Chen:2018dbv}, the baryon acoustic oscillation (BAO) data from the 6dF Galaxy Survey (6dFGS), the Main Galaxy Sample of Data Release 7 of Sloan Digital Sky Survey (SDSS-MGS), and the Data Release 12 galaxy sample of Baryon Oscillation Spectroscopic Survey (BOSS-DR12) \cite{Beutler:2011hx,Ross:2014qpa,Alam:2016hwk}, and the type Ia supernova (SN) data from the Pantheon compilation \cite{Scolnic:2017caz}. We constrain the various dark energy models by using these cosmological data and use the Markov-chain Monte Carlo (MCMC) approach \cite{Lewis:2002ah} to infer the posterior distributions of parameters. The best-fit results obtained are used to serve for the fiducial cosmological models in the simulation of GW data. Here we note that we do not use one unique fiducial model in this work, but instead for each specific dark energy model we wish to analyze we actually use  this model itself as a fiducial model in the analysis for it. For example, in the analysis of the $w$CDM model, we use the $w$CDM model itself as the fiducial model to simulate the GW data. We simulate 1000 GW standard siren data based on the 10-year observation of the Einstein Telescope (ET) \cite{ET}, which is a third-generation ground-based GW detector.\footnote{It should be mentioned that, in addition to ET in Europe, there is another leading proposal for the design of the third-generation GW detector, i.e., the Cosmic Explorer (CE) in the United States \cite{Evans:2016mbw,Dwyer:2014fpa}. The design of CE is rather different from that of ET. For the scientific potential of CE, we refer the reader to Ref.~\cite{Zhao:2017cbb}. In this paper, we only focus on the discussion of ET.} In order to avoid the discrepancy of the central values from the GW data and the EM data in the parameter planes, we omit the step of a random gaussian sampling for the fiducial cosmology in the simulation. Therefore, this work can overcome the drawbacks in the investigation of Ref.~\cite{Zhang:2018byx}.

It should be pointed out that the GW observations not only can be used to constrain dark energy models, but also can exert significant influences on the studies of theories of modified gravity (MG). For example, the measurement of the propagation speed of GWs using the observation of GW170817 \cite{Monitor:2017mdv} has immediately been used to exclude some MG models \cite{Baker:2017hug,Creminelli:2017sry,Sakstein:2017xjx,Ezquiaga:2017ekz}. The impacts of the future GW observations on MG models have been recently intensively discussed, which can be found in, e.g., Refs.~\cite{Will:1994fb,Liu:2017xef,Berti:2018cxi,Zhao:2018gwk,Liu:2018sia}. But in this paper we confine our discussions only to the dark energy cosmological models.

This work is organized as follows. In Sec.~\ref{sec2}, we describe the method to simulate the GW standard siren data from the ET and give a briefly introduction of these typical dark energy models, we also introduce the conventional cosmological probes based on the EM observations and the method to constrain cosmological parameters. In Sec.~\ref{sec3}, we report and discuss the constraint results. In Sec.~\ref{sec4}, we give the conclusion of this work.

\section{Method and data}\label{sec2}

\subsection{Method of simulating GW data}

For a  Friedmann-Robertson-Walker universe, the line element reads
\begin{equation}
d{s^2} = -d{t^2} + {a^2}(t)\left[\frac{{d{r^2}}}{{1 - K{r^2}}} + {r^2}(d{\theta ^2} + {\sin ^2}\theta d{\phi ^2})\right],
\label{equa:ds}
\end{equation}
where $t$ is the cosmic time, $a(t)$ is the scale factor, and $K = +1,-1,0$ corresponds to a closed, open, and flat universe, respectively. We set $G=c=1$ and $K = 0$ throughout this paper. Then the luminosity distance $d_L$ can be written as
\begin{equation}
{d_L} (z)= \frac{{(1 + z)}}{{H_0}}\int_0^z {\frac{{dz'}}{{E(z')}}},
\label{equa:dl}
\end{equation}
where $E(z)\equiv H(z)/H_0$ (the Hubble constant $H_{0}= 100h$ $\rm km~s^{-1}~Mpc^{-1}$). For the different dark energy models, the concrete expressions of $E(z)$ can be found in the next subsection (see also Ref.~\cite{Xu:2016grp}).

The first step is to simulate the redshift distribution of the sources. We constrain cosmological parameters by simulating many catalogues of the mergers of BNS or of a neutron star and a black hole (BHNS). Following Ref.~\cite{Li:2013lza}, the neutron star (NS) mass distribution is taken to be uniform in the interval [1,2] $M_\odot$, and the black hole (BH) mass distribution is taken to be uniform between [3,10] $M_\odot$, where $M_\odot$ is the solar mass. The number ratio between BHNS and BNS events is taken to be 0.03, as predicted for the Advanced LIGO-Virgo network \cite{Abadie:2010px}. Following Refs.~\cite{Cai:2016sby,Zhao:2010sz,Wang:2018lun}, the redshift distribution of the sources takes the form
\begin{equation}
P(z)\propto \frac{4\pi d_C^2(z)R(z)}{H(z)(1+z)},
\label{equa:pz}
\end{equation}
where $d_C$ is the comoving distance, which is defined as $d_C(z)\equiv\int_0^z {1/H(z')dz'}$, and $R(z)$ describes the time evolution of the burst rate and takes the form~\cite{Schneider:2000sg,Cutler:2009qv}
\begin{equation}
R(z)=\begin{cases}
1+2z, & z\leq 1 \\
\frac{3}{4}(5-z), & 1<z<5 \\
0, & z\geq 5.
\end{cases}
\label{equa:rz}
\end{equation}


Following Ref.~\cite{Cai:2016sby}, the strain in GW interferometers can be written as
\begin{equation}
h(t)=F_+(\theta, \phi, \psi)h_+(t)+F_\times(\theta, \phi, \psi)h_\times(t),
\end{equation}
where $\psi$ is the polarization angle and ($\theta$,$\phi$) are angles describing the location of the source relative to the detector, and $F_{+}$ and $F_{\times}$ are the antenna pattern functions of the ET \cite{Zhao:2010sz}, with the forms written as
\begin{align}
F_+^{(1)}(\theta, \phi, \psi)=&~~\frac{{\sqrt 3 }}{2}[\frac{1}{2}(1 + {\cos ^2}(\theta ))\cos (2\phi )\cos (2\psi ) \nonumber\\
                              &~~- \cos (\theta )\sin (2\phi )\sin (2\psi )],\nonumber\\
F_\times^{(1)}(\theta, \phi, \psi)=&~~\frac{{\sqrt 3 }}{2}[\frac{1}{2}(1 + {\cos ^2}(\theta ))\cos (2\phi )\sin (2\psi ) \nonumber\\
                              &~~+ \cos (\theta )\sin (2\phi )\cos (2\psi )].
\label{equa:F}
\end{align}
The three interferometers have $60^\circ$ with each other, so the other two antenna pattern functions are $F_{+,\times}^{(2)}(\theta, \phi, \psi)=F_{+,\times}^{(1)}(\theta, \phi+2\pi/3, \psi)$ and $F_{+,\times}^{(3)}(\theta, \phi, \psi)=F_{+,\times}^{(1)}(\theta, \phi+4\pi/3, \psi)$, respectively.

Following Refs.~\cite{Zhao:2010sz,Li:2013lza}, we could know the Fourier transform $\mathcal{H}(f)$ of the time domain waveform $h(t)$,
\begin{align}
\mathcal{H}(f)=\mathcal{A}f^{-7/6}\exp[i(2\pi ft_0-\pi/4+2\psi(f/2)-\varphi_{(2.0)})],
\label{equa:hf}
\end{align}
where $\mathcal{A}$ is the Fourier amplitude that is written as
\begin{align}
\mathcal{A}=&~~\frac{1}{d_L}\sqrt{F_+^2(1+\cos^2(\iota))^2+4F_\times^2\cos^2(\iota)}\nonumber\\
            &~~\times \sqrt{5\pi/96}\pi^{-7/6}\mathcal{M}_c^{5/6},
\label{equa:A}
\end{align}
where $\mathcal{M}_c=M \eta^{3/5}$ is called ``chirp mass", $M=m_1+m_2$ is the total mass of coalescing binary with component masses $m_1$ and $m_2$, and $\eta=m_1 m_2/M^2$ is the symmetric mass ratio. Here, we need to state that the observed mass $\mathcal{M}_{c,\rm obs}=(1+z)\mathcal{M}_{c,\rm phys}$. $\mathcal{M}_c$ in Eq.~(\ref{equa:A}) represents the observed mass. $\iota$ is the angle of inclination of the binary's orbital angular momentum with the line of sight. The definitions of the functions $\psi$ and $\varphi_{(2.0)}$ can refer to Refs.~\cite{Zhao:2010sz,Li:2013lza}. Since it is expected that the short gamma ray bursts (SGRBs) are expected to be strongly beamed \cite{Abdo:2009zza,Nakar:2005bs,Rezzolla:2011da}, the coincidence observations of SGRBs imply that the binaries are orientated nearly face on (i.e., $\iota\simeq 0$) and the maximal inclination is about $\iota=20^\circ$. Actually, averaging the Fisher matrix over the inclination $\iota$ and the polarization $\psi$ with the constraint $\iota<20^\circ$ is approximately the same as taking $\iota=0$ in the simulation \cite{Li:2013lza}. Therefore, we can take $\iota=0$ in the process of simulating GW sources.

The performance of a GW detector is characterized by the one-side noise \textit{power spectral density} $S_h(f)$ (PSD). We take the noise PSD of the ET to be the same as in Ref.~\cite{Zhao:2010sz}. The combined signal-to-noise (SNR) for the network of three independent interferometers is
\begin{equation}
\rho=\sqrt{\sum\limits_{i=1}^{3}(\rho^{(i)})^2},
\label{euqa:rho}
\end{equation}
where $\rho^{(i)}=\sqrt{\left\langle \mathcal{H}^{(i)},\mathcal{H}^{(i)}\right\rangle}$. The inner product is defined as \begin{equation}
\left\langle{a,b}\right\rangle=4\int_{f_{\rm lower}}^{f_{\rm upper}}\frac{\tilde a(f)\tilde b^\ast(f)+\tilde a^\ast(f)\tilde b(f)}{2}\frac{df}{S_h(f)},
\label{eq:product}
\end{equation}
where $\tilde a(f)$ and $\tilde b(f)$ are the Fourier transforms of the functions $a(t)$ and $b(t)$. For more details, see Ref.~\cite{Cai:2016sby}.


Using the Fisher information matrix, we can get the instrumental error of $d_{L}$,
\begin{align}
\sigma_{d_L}^{\rm inst}\simeq \sqrt{\left\langle\frac{\partial \mathcal H}{\partial d_L},\frac{\partial \mathcal H}{\partial d_L}\right\rangle^{-1}}.
\end{align}
Because $\mathcal H \propto d_L^{-1}$, we obtain $\sigma_{d_L}^{\rm inst}\simeq d_L/\rho$. Considering the effect from the inclination angle $\iota$ (between $\iota=0$ and $\iota=90^\circ$), we add a factor 2 in front of the error. Therefore, the true instrumental error of $d_{L}$ is written as
\begin{equation}
\sigma_{d_L}^{\rm inst}\simeq \frac{2d_L}{\rho}.
\label{sigmainst}
\end{equation}
Following Ref.~\cite{Cai:2016sby}, we can get the additional error from weak lensing, $\sigma_{d_L}^{\rm lens}$ = $0.05z d_L$. Thus, actually, the total error of $d_{L}$ is
\begin{align}
\sigma_{d_L}&~~=\sqrt{(\sigma_{d_L}^{\rm inst})^2+(\sigma_{d_L}^{\rm lens})^2} \nonumber\\
            &~~=\sqrt{\left(\frac{2d_L}{\rho}\right)^2+(0.05z d_L)^2}.
\label{sigmadl}
\end{align}

So far, we could get all the information of GW events, including $z$, $d_{L}$, and $\sigma_{d_{L}}$. Therefore, we could simulate 1000 GW events expected to be detected by ET in its 10-yr observation.

\begin{figure*}[!htp]
\includegraphics[width=12cm]{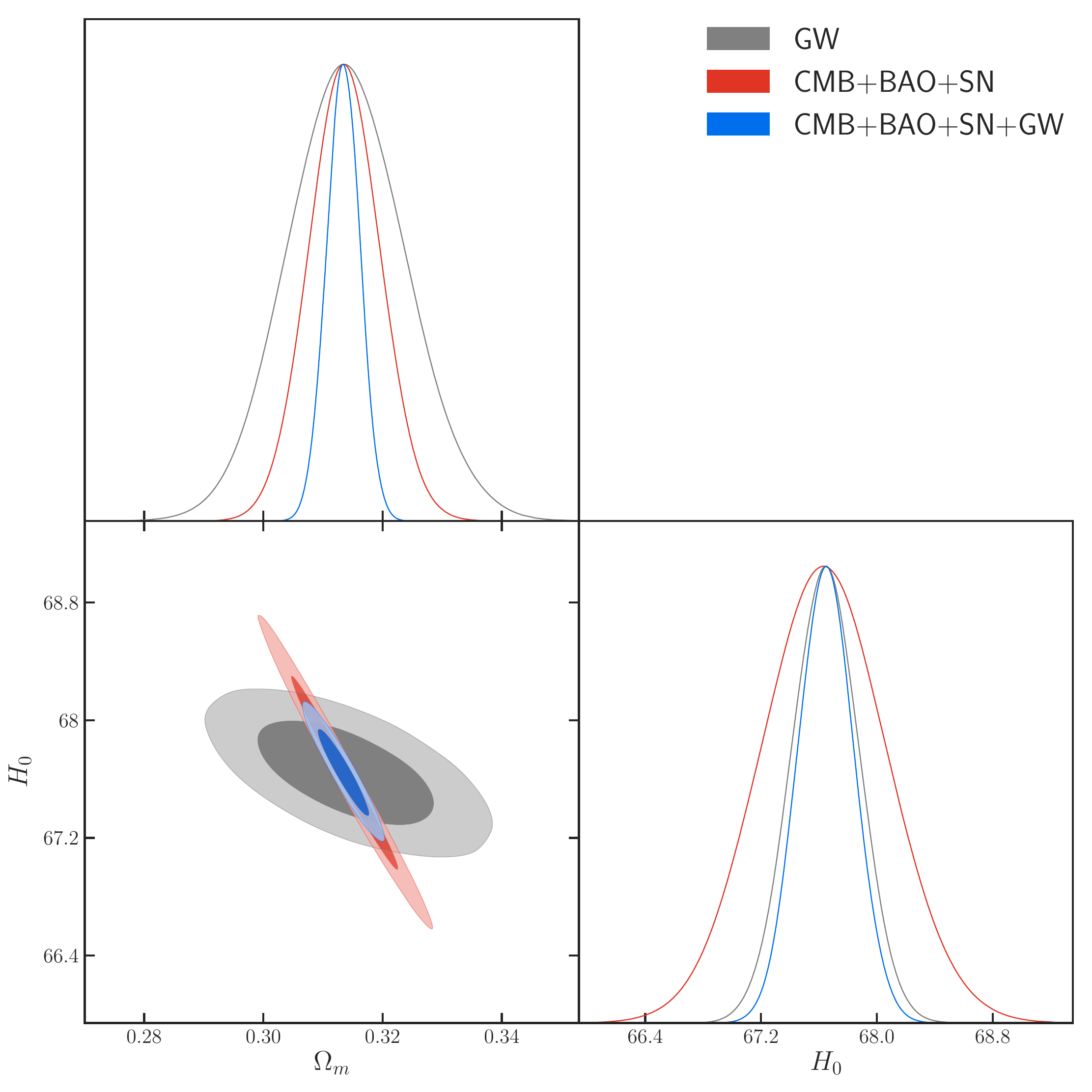}
\centering
 \caption{\label{fig1}  Constraints (68.3\% and 95.4\% confidence level) on the $\Lambda$CDM model by using the GW, CMB+BAO+SN, and CMB+BAO+SN+GW data. Here, $H_0$ is in units of km s$^{-1}$ Mpc$^{-1}$, and this is the same for all the figures in this paper.}
\end{figure*}

\begin{figure*}[!htp]
\includegraphics[width=12cm]{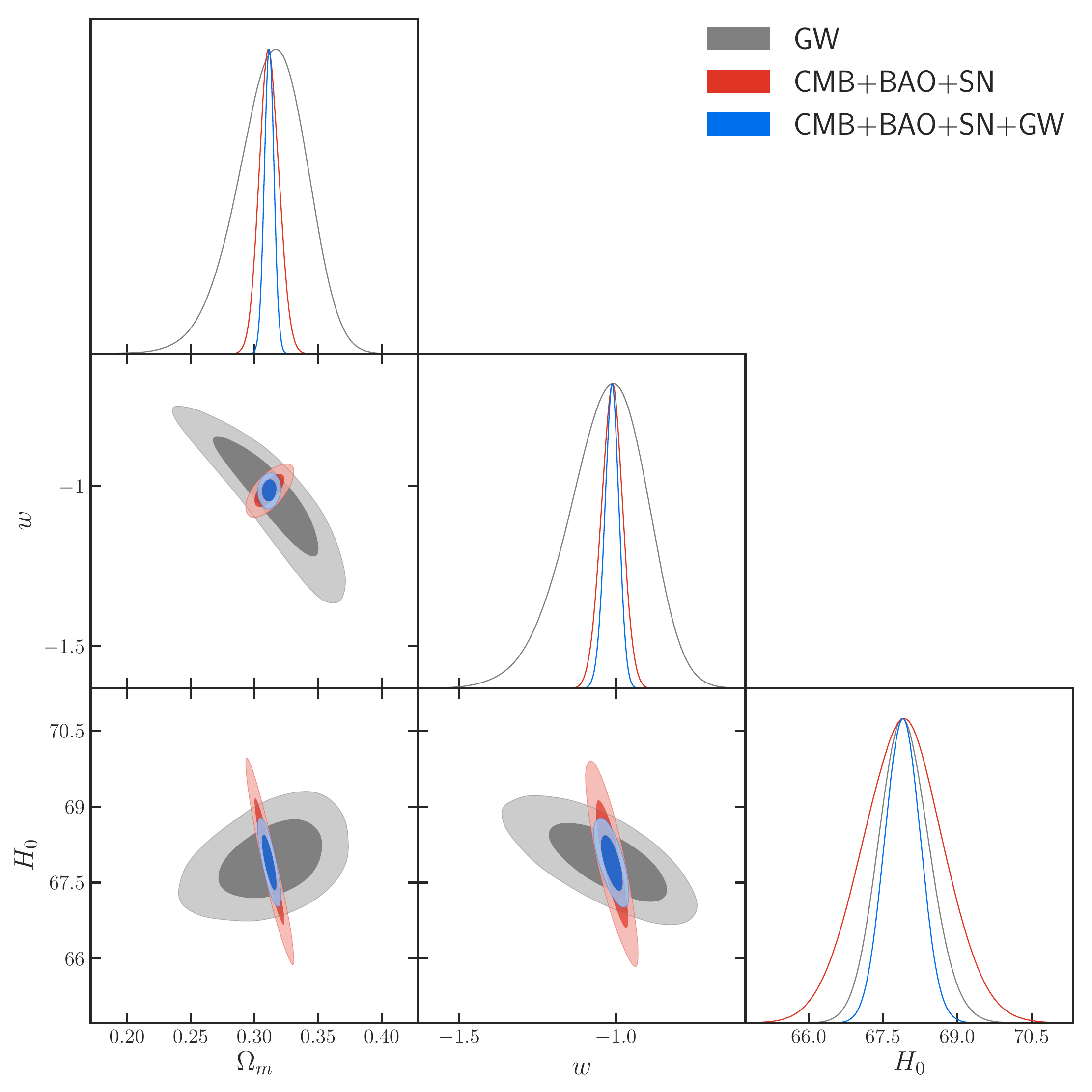}
\centering
 \caption{\label{fig2} Constraints (68.3\% and 95.4\% confidence level) on the $w$CDM model by using the GW, CMB+BAO+SN, and CMB+BAO+SN+GW data.}
\end{figure*}

\begin{table*}[htbp]
\small
\setlength\tabcolsep{2.8pt}
\renewcommand{\arraystretch}{1.5}
\centering
\begin{tabular}{c|cccc}
\hline
\hline
Model & Parameter  &  CMB+BAO+SN  &  GW & CMB+BAO+SN+GW    \\
\hline
\multirow{2}{*}{$ \Lambda $CDM} & $ \Omega_{\rm{m}} $ & $ 0.3136^{+0.0060}_{-0.0060} $ & $ 0.3140^{+0.0098}_{-0.0098} $ & $ 0.3134^{+0.0028}_{-0.0028} $ \\
& $H_0$ & $ 67.64^{+0.44}_{-0.44} $ & $ 67.64^{+0.23}_{-0.23} $ & $ 67.65^{+0.19}_{-0.19} $ \\
\hline
\multirow{3}{*}{$ w $CDM} & $ \Omega_{\rm{m}} $ & $ 0.3117^{+0.0078}_{-0.0078} $ & $ 0.312^{+0.029}_{-0.025} $ & $ 0.3116^{+0.0044}_{-0.0044} $ \\
& $H_0$ & $ 67.9^{+0.83}_{-0.83} $ & $ 67.96^{+0.69}_{-0.69} $ & $ 67.90^{+0.46}_{-0.46} $ \\
& $ w $ & $ -1.013^{+0.034}_{-0.034} $ & $ -1.03^{+0.14}_{-0.11} $ & $ -1.013^{+0.026}_{-0.026} $ \\
\hline
\multirow{4}{*}{CPL} & $ \Omega_{\rm{m}} $ & $ 0.3122^{+0.0079}_{-0.0079} $ & $ 0.327^{+0.091}_{-0.043} $ & $ 0.3118^{+0.0068}_{-0.0068} $ \\
& $H_0$ & $ 67.91^{+0.83}_{-0.83} $ & $ 67.2^{+2.6}_{-2.3} $ & $ 67.93^{+0.67}_{-0.67} $ \\
& $ w_0 $ & $ -0.993^{+0.083}_{-0.083} $ & $ -0.76^{+0.28}_{-0.53} $ & $ -0.997^{+0.073}_{-0.073} $ \\
& $ w_a $ & $ -0.10^{+0.36}_{-0.27} $ & $ -1.8^{+3.4}_{-1.2} $ & $ 0.08^{+0.27}_{-0.23} $ \\
\hline
\multirow{3}{*}{$\alpha$DE}& $ \Omega_{\rm{m}} $ & $ 0.3114^{+0.0076}_{-0.0076} $ & $ 0.307^{+0.034}_{-0.021} $ & $ 0.3114^{+0.0043}_{-0.0043} $ \\
& $H_0$ & $ 67.97^{+0.82}_{-0.82} $ & $ 67.95^{+0.61}_{-0.61} $ & $ 67.95^{+0.45}_{-0.45} $ \\
& $ \alpha $ & $ -0.07^{+0.15}_{-0.12} $ & $ -0.07^{+0.49}_{-0.60} $ & $ -0.064^{+0.11}_{-0.091} $ \\
\hline
\multirow{3}{*}{GCG}& $ A_{\rm s} $ & $ 0.728^{+0.022}_{-0.022} $ & $ 0.733^{+0.051}_{-0.051} $ & $ 0.729^{+0.019}_{-0.019} $ \\
& $ \beta $ & $ 0.009^{+0.075}_{-0.075} $ & $ 0.05^{+0.20}_{-0.27} $ & $ 0.009^{+0.059}_{-0.067} $ \\
& $ H_0 $ & $ 67.96^{+0.42}_{-0.38} $ & $ 68.00^{+0.65}_{-0.65} $ & $ 67.97^{+0.34}_{-0.34} $ \\
\hline
\multirow{4}{*}{NGCG}& $ w $ & $ -1.002^{+0.045}_{-0.045} $ & $ -1.32^{+0.53}_{-0.18} $ & $ -1.003^{+0.027}_{-0.027} $ \\
& $H_0$ & $ 67.78^{+0.87}_{-0.87} $ & $ 67.81^{+0.26}_{-0.26} $ & $ 67.79^{+0.23}_{-0.23} $ \\
& $ \beta $ & $ -0.0029^{+0.0097}_{-0.0097} $ & $ -0.22^{+0.26}_{-0.49} $ & $ -0.0026^{+0.0087}_{-0.0087} $ \\
& $ \Omega_{\rm de} $ & $ 0.6879^{+0.0078}_{-0.0078} $ & $ 0.58^{+0.21}_{-0.11} $ & $ 0.6880^{+0.0032}_{-0.0032} $ \\
\hline
\hline
\end{tabular}
\caption{\label{tab1} Constraint results of the $\Lambda$CDM, $w$CDM, CPL, $\alpha$DE, GCG, and NGCG models using the CMB+BAO+SN, GW, and CMB+BAO+SN+GW data. Here, $H_0$ is in units of km s$^{-1}$ Mpc$^{-1}$, and this is the same for all the tables in this paper.}
\centering
\renewcommand{\arraystretch}{1.5}
\end{table*}

\begin{figure*}[!htp]
\includegraphics[width=12cm]{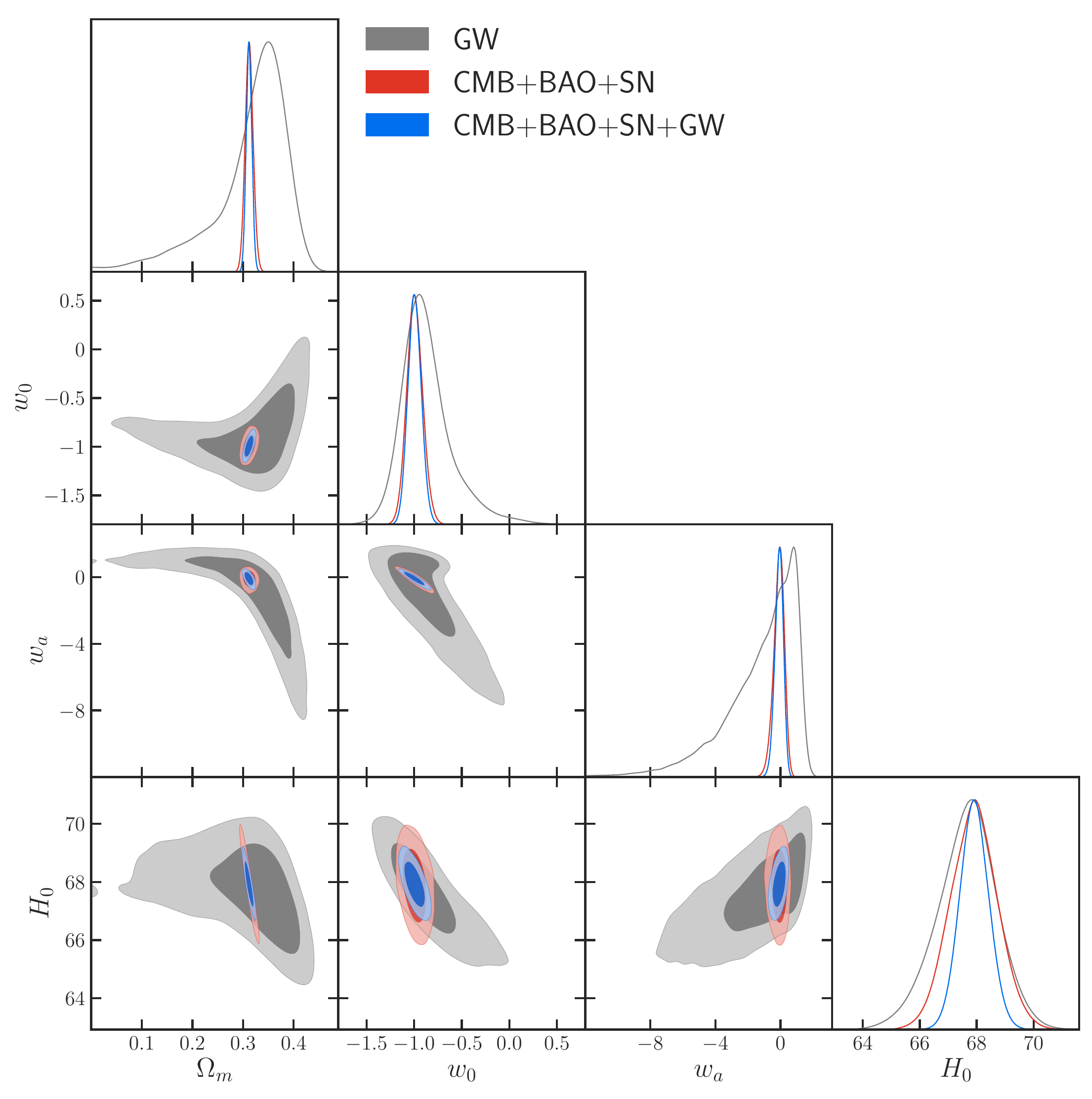}
\centering
 \caption{\label{fig3} Constraints (68.3\% and 95.4\% confidence level) on the CPL model by using the GW, CMB+BAO+SN, and CMB+BAO+SN+GW data.}
\end{figure*}

\begin{figure*}[!htp]
\includegraphics[width=12cm]{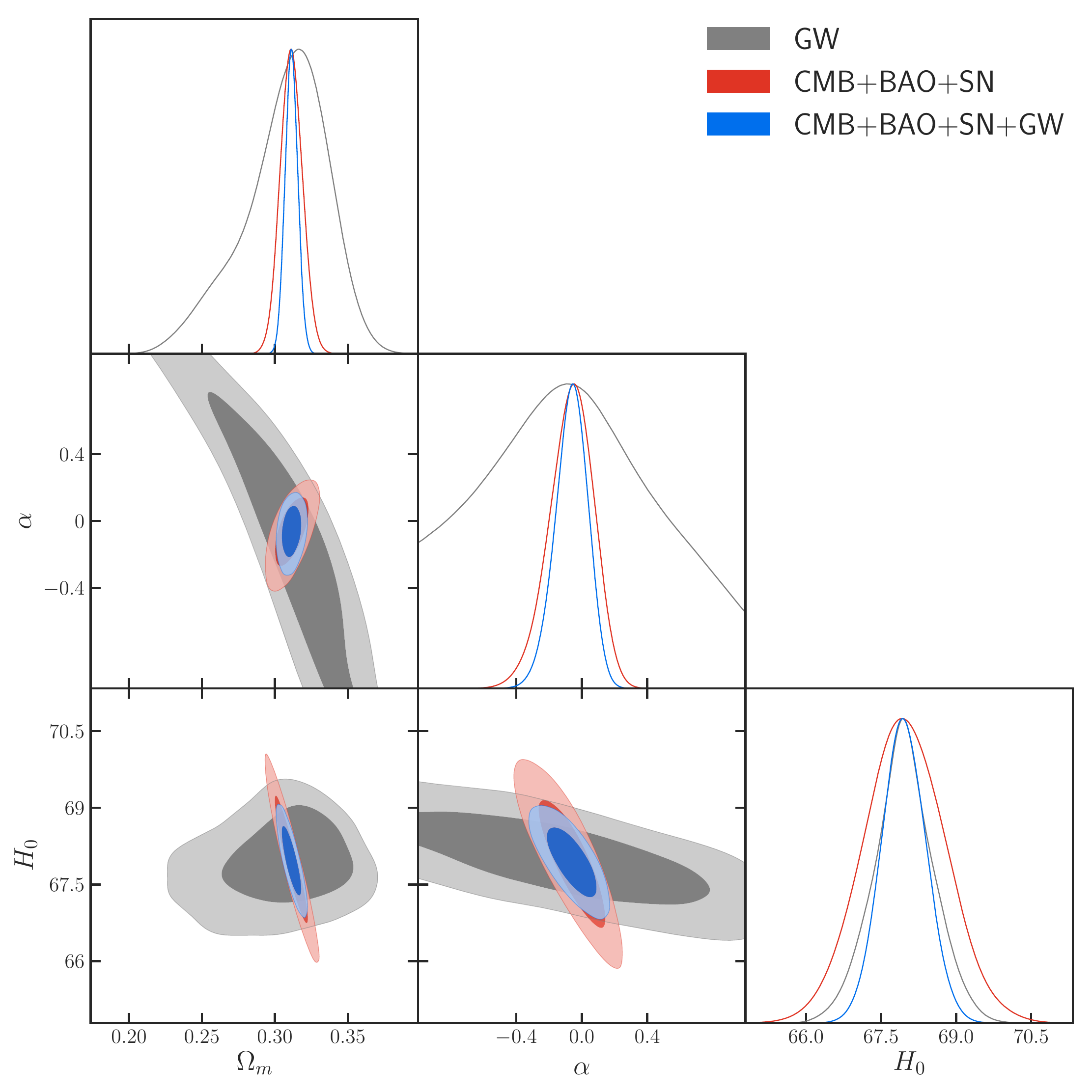}
\centering
 \caption{\label{fig4} Constraints (68.3\% and 95.4\% confidence level) on the $\alpha$DE model by using the GW, CMB+BAO+SN, and CMB+BAO+SN+GW data.}
\end{figure*}

\subsection{Dark energy models}

In this subsection, we give a briefly description of the dark energy models, including $\Lambda$CDM model, $w$CDM model, CPL model, $\alpha$DE model, GCG model and NGCG model.

\begin{itemize}
\item $\Lambda$CDM model: Nowadays, the cosmological constant $\Lambda$ is the most promising candidate for dark energy accounting for the current acceleration of the universe. The cosmological model with $\Lambda$ and cold dark matter (CDM) is called the $\Lambda$CDM model. The equation of state (EoS) of the cosmological constant (or the vacuum energy density) is $w = -1$, so we have
\begin{equation}
E^2(z)=\Omega_{\rm{m}}(1+z)^{3}+\Omega_{\rm{r}}(1+z)^{4}+(1-\Omega_{\rm{m}}-\Omega_{\rm{r}}).
\end{equation}

\item $w$CDM model: In this model, the EoS of dark energy is $w = \rm{constant}$. This model is the simplest case for describing a dynamical dark energy. The form of $E(z)$ of this model is written as
\begin{equation}
E^2(z)=\Omega_{\rm{m}}(1+z)^{3}+\Omega_{\rm{r}}(1+z)^{4}+(1-\Omega_{\rm{m}}-\Omega_{\rm{r}})(1+z)^{3(1+w)}.
\end{equation}

\item CPL model: This model is a parametrization model of dark energy for generally describing the evolution of $w(z)$, which is usually also called the $w_{0}w_{a}$CDM model. The form of $w(z)$ in this model is written as
\begin{equation}
w(z)=w_{\rm{0}}+w_{\rm{a}}\frac{z}{1+z},
\end{equation}
where $w_{0}$ and $w_{\rm a}$ are free parameters. In this model, we have
\begin{equation}
\begin{aligned}
E^2(z)&=\Omega_{\rm{m}}(1+z)^{3}+\Omega_{\rm{r}}(1+z)^{4}\\
&+(1-\Omega_{\rm{m}}-\Omega_{\rm{r}})(1+z)^{3(1+w_{\rm{0}}+w_{\rm{a}})}\exp\left(-\frac{3w_{\rm{a}}z}{1+z}\right).
 \end{aligned}
\end{equation}

\item $\alpha$DE model: The Dvali-Gabadadze-Porrati (DGP) braneworld model \cite{a:2000} is a well-known example of the modified gravity. As a phenomenological extension of the DGP model, the $\alpha$DE model can fit the observational data much better, in which the Friedmann equation is modified as
\begin{equation}
3M^{2}_{\rm{pl}}\left(H^{2}-\frac{H^{\alpha}}{r^{2-\alpha}_{\rm{c}}}\right)=\rho_{\rm{m}}(1+z)^{3}+\rho_{\rm{r}}(1+z)^{4},
\end{equation}
where $\alpha$ is a phenomenological parameter and $r_{\rm{c}}=(1-\Omega_{\rm{m}}-\Omega_{\rm{r}})^{1/(\alpha-2)}H^{-1}_{0}$. In this model, $E(z)$ is derived by solving the following equation
\begin{equation}
E^2(z)=\Omega_{\rm{m}}(1+z)^{3}+\Omega_{\rm{r}}(1+z)^{4}+E^{\alpha}(z)(1-\Omega_{\rm{m}}-\Omega_{\rm{r}}).
\end{equation}
Obviously, the model with $\alpha=1$ reduces to the DGP model \cite{a:2000} and the model with $\alpha=0$ reduces to the $\Lambda$CDM model.

\item GCG model: The Chaplygin gas model \cite{Kamenshchik:2001cp}, which is generally viewed as arising from the $d$-brane theory, can describe the cosmic acceleration and provide a unification scheme for vacuum energy and cold dark matter. The original Chaplygin gas model has been excluded by the current observations \cite{c:2003}, therefore here we consider GCG model \cite{MCB:2002}. The EoS of the GCG fluid is
\begin{equation}
p_{\rm{gcg}}=-\frac{A}{\rho^{\beta}_{\rm{gcg}}},
\end{equation}
where $A$ is a positive constant and $\beta$ is a free parameter. The energy density of the GCG fluid can be derived,
\begin{equation}
\rho_{\rm{gcg}}(a)=\rho_{\rm{gcg}0}\left(A_{\rm{s}}+\frac{1-A_{\rm{s}}}{a^{3(1+\beta)}}\right)^{\frac{1}{1+\beta}},
\end{equation}
where $A_{\rm{s}}\equiv A/\rho^{1+\beta}_{\rm{gcg}0}$ is a dimensionless parameter. Thus we can derive the form of $E(z)$ for this model as
\begin{equation}
E^2(z)=\Omega_{\rm{b}}(1+z)^{3}+\Omega_{\rm{r}}(1+z)^{4}+(1-\Omega_{\rm{b}}-\Omega_{\rm{r}})\left(A_{\rm{s}}+(1-A_{\rm{s}})(1+z)^{3(1+\beta)}\right)^{1\over 1+\beta},
\end{equation}
where $\Omega_{\rm b}$ is the present-day density of baryon matter. Obviously, the GCG model with $\beta=0$ reduces to the $\Lambda$CDM model and with $\beta=1$ reduces to the original Chaplygin gas model.

\item NGCG model: Inspired by the GCG model, the NGCG model is proposed in Ref.~\cite{Zhang:2004gc}. The EoS of the NGCG fluid can be written as
\begin{equation}
p_{\rm{ngcg}}=-\frac{\tilde{A}(a)}{\rho^{\beta}_{\rm{ngcg}}},
\end{equation}
where $\tilde{A}(a)$ is a function of the scale factor $a$ and $\beta$ is a free parameter. The energy density of the NGCG fluid can be written as
\begin{equation}
\rho_{\rm{ngcg}}=\left[Aa^{-3(1+w)(1+\beta)}+Ba^{-3(1+\beta)}\right]^{\frac{1}{1+\beta}},\label{ngcg1}
\end{equation}
where $A$ and $B$ are positive constants. The form of the function $\tilde{A}(a)$ is
\begin{equation}
\tilde{A}(a)=-wAa^{-3(1+w)(1+\beta)}.
\end{equation}
The form of $E(z)$ of this model is given by
\begin{equation}
\begin{aligned}
E^2(z)&=\Omega_{\rm{b}}(1+z)^{3}+\Omega_{\rm{r}}(1+z)^{4}+(1-\Omega_{\rm{b}}-\Omega_{\rm{r}})(1+z)^{3}\\
&\times \left[1-\frac{\Omega_{\rm{de}}}{1-\Omega_{\rm{b}}-\Omega_{\rm{r}}}\left(1-(1+z)^{3w(1+\beta)}\right)\right]^{1\over 1+\beta}.\label{ngcg}
\end{aligned}
\end{equation}
Obviously, the NGCG model with $w=-1$ reduces to the GCG model.

\end{itemize}

\subsection{Conventional cosmological probes}

In this paper, we use the conventional cosmological probes to constrain the cosmological parameters of various dark energy models. Using the fitting results, we can analyze the different models and simulate the GW data. In the following, we give a brief description of the data used in this paper.

\begin{itemize}
\item The CMB data: We use the ``Planck distance priors" from the Planck 2018 results, including the shift parameter $R=1.750235$, the ``acoustic scale" $\ell_{\rm{A}}=301.4707$, and the baryon density $\omega_{\rm{b}}\equiv\Omega_{\rm{b}}h^{2}=0.02235976$. More details could refer to Ref.~\cite{Aghanim:2018eyx}.

\item The BAO data: The BAO observation can be used to measure the angular diameter distance and the expansion rate of the universe. We take the measurements from 6dFGS at $z_{\rm eff} = 0.106$ \cite{Beutler:2011hx}, SDSS-MGS at $z_{\rm eff} = 0.15$ \cite{Ross:2014qpa}, and BOSS-DR12 at the effective redshifts of 0.38, 0.51, and 0.61 \cite{Alam:2016hwk}.

\item The SN data: We consider the latest ``Pantheon" sample \cite{Scolnic:2017caz}. The total number of SN is 1048 in the redshift range of $z\in[0.01, 2.3]$.

\end{itemize}

\subsection{Method of constraining parameters}

In order to constrain cosmological parameters, we use the MCMC method to infer their posterior probability distributions. For the combination of CMB, BAO, and SN data, the total $\chi^2_{\rm tot}$ is
\begin{equation}
\chi^2_{\rm tot} = \chi^2_{\rm CMB} + \chi^2 _{\rm BAO} +\chi^2_{\rm SN}.
\label{chi1}
\end{equation}

In this paper, we also use the 1000 simulated GW data points in the cosmological fit. For the GW data, its $\chi^2$ can be written as
\begin{align}
\chi_{\rm GW}^2=\sum\limits_{i=1}^{1000}\left[\frac{\bar{d}_L^i-d_L(\bar{z}_i;\vec{\Omega})}{\bar{\sigma}_{d_L}^i}\right]^2,
\label{chi2}
\end{align}
where $\bar{z}_i$, $\bar{d}_L^i$, and $\bar{\sigma}_{d_L}^i$ are the $i$th redshift, luminosity distance, and error of luminosity distance, respectively. $\vec{\Omega}$ denotes a set of cosmological parameters.

If we consider the combination of the conventional cosmological EM observations and the GW standard siren observation, the total $\chi^2_{\rm tot}$ becomes
\begin{equation}
\chi^2_{\rm tot} = \chi^2_{\rm CMB} + \chi^2 _{\rm BAO} +\chi^2_{\rm SN} +\chi^2_{\rm GW}.
\label{chi3}
\end{equation}

\section{Results}\label{sec3}

\begin{figure*}[!htp]
\includegraphics[width=12cm]{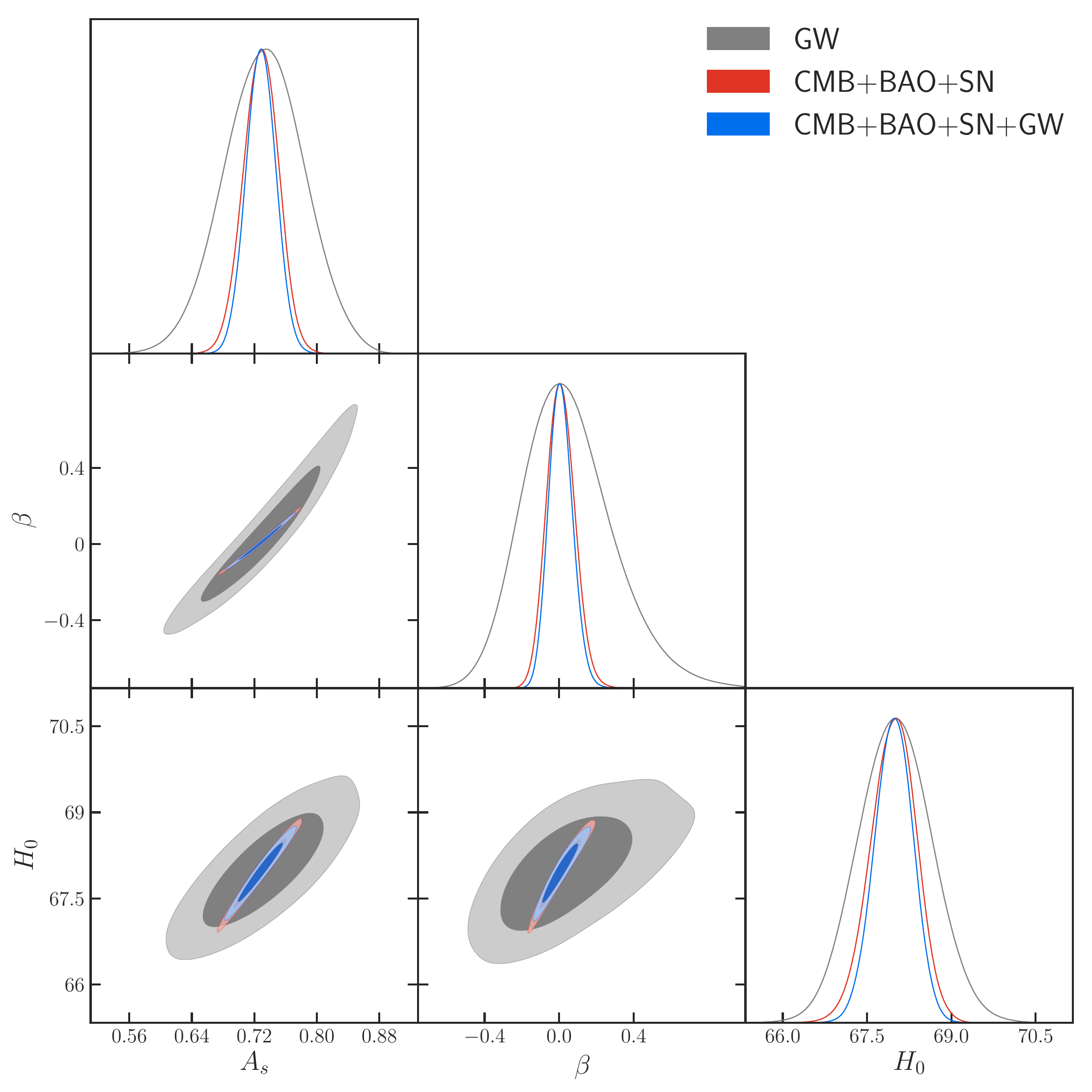}
\centering
 \caption{\label{fig5} Constraints (68.3\% and 95.4\% confidence level) on the GCG model by using the GW, CMB+BAO+SN, and CMB+BAO+SN+GW data.}
\end{figure*}

\begin{figure*}[!htp]
\includegraphics[width=12cm]{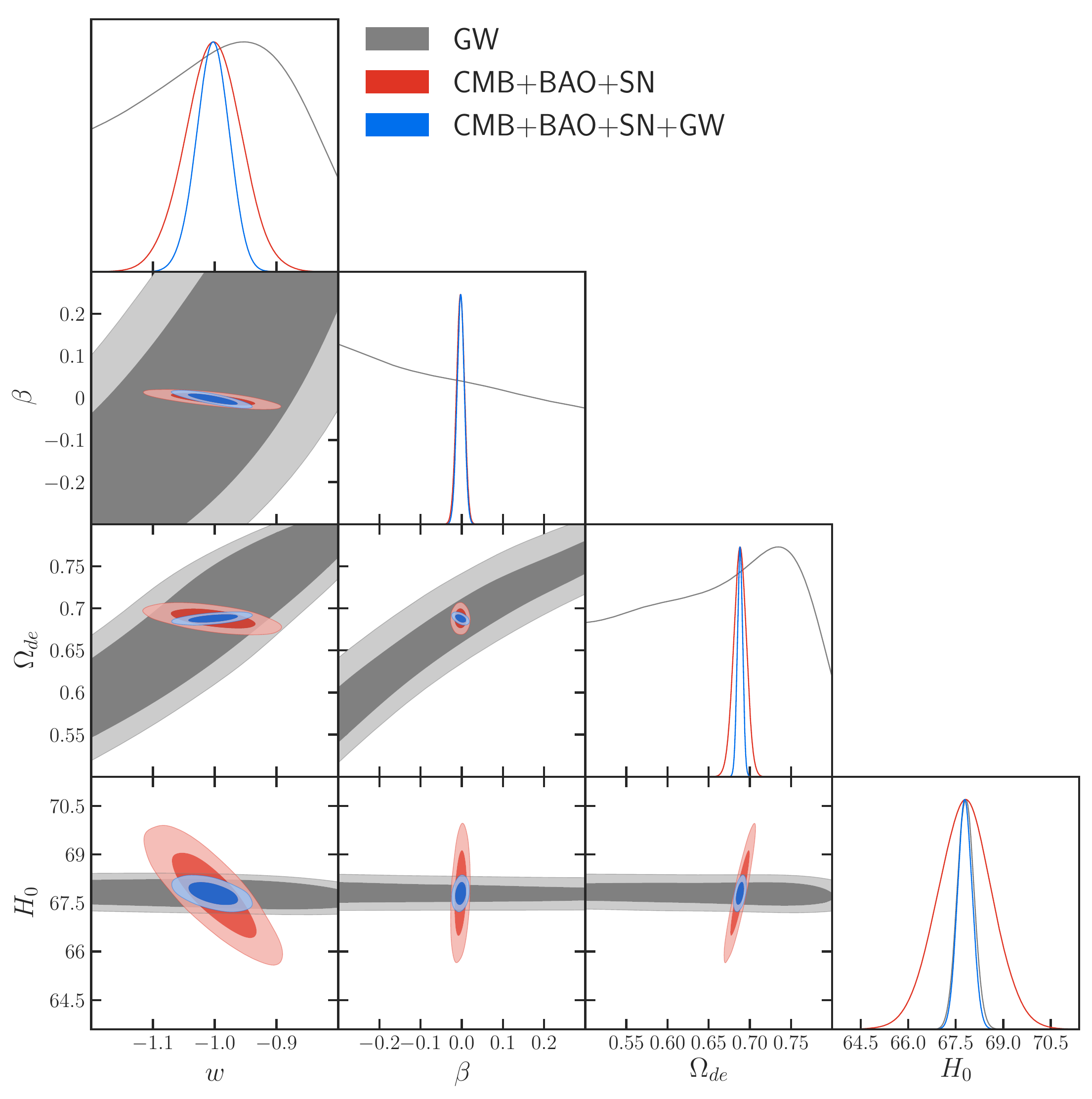}
\centering
 \caption{\label{fig6} Constraints (68.3\% and 95.4\% confidence level) on the NGCG model by using the GW, CMB+BAO+SN, and CMB+BAO+SN+GW data.}
\end{figure*}

\begin{table*}[htbp]
\small
\setlength\tabcolsep{2.8pt}
\renewcommand{\arraystretch}{1.5}
\centering
\begin{tabular}{cccccccccccc}
\\
\hline\hline &\multicolumn{2}{c}{CMB+BAO+SN}&& \multicolumn{2}{c}{GW}&& \multicolumn{2}{c}{CMB+BAO+SN+GW} \\
 \cline{2-3}\cline{5-6}\cline{8-9}
 Error  & $\Lambda$CDM & $w$CDM & & $\Lambda$CDM & $w$CDM   & &$\Lambda$CDM & $w$CDM \\
\hline
$\sigma(\Omega_{\rm m})$
                   & $0.0060$
                   & $0.0078$&
                   & $0.0098$
                   & $0.0270$
                   && $0.0028$
                   & $0.0037$\\

$\sigma(H_0)$
                    & $0.4400$
                   & $0.8300$&
                   & $0.2300$
                   & $0.5200$
                   && $0.1900$
                   & $0.3600$\\

$\sigma(w)$
                    & $-$
                   & $0.0340$&
                   & $-$
                   & $0.1250$
                   && $-$
                   & $0.0230$\\
                      \cline{2-3}\cline{5-6}\cline{8-9}

Error  & CPL & $\alpha$DE && CPL & $\alpha$DE && CPL & $\alpha$DE \\ \hline

$\sigma(\Omega_{\rm m})$
                   & $0.0079$
                   & $0.0076$&
                   & $0.0560$
                   & $0.0275$
                   && $0.0056$
                   & $0.0043$\\

$\sigma(H_0)$
                  & $0.8300$
                   & $0.8200$&
                   & $1.0250$
                   & $0.6100$
                  & & $0.5100$
                   & $0.4500$\\

$\sigma(\alpha)$
                   & $-$
                   & $0.1350$&
                   & $-$
                   & $0.5450$
                   && $-$
                   & $0.1005$\\

$\sigma(w_0)$
                   & $0.0830$
                   & $-$&
                   & $0.2300$
                   & $-$
                   && $0.0700$
                   & $-$\\

$\sigma(w_a)$
                   & $0.3150$
                   & $-$&
                   & $1.6850$
                   & $-$
                   && $0.2550$
                   & $-$\\
\hline\hline
 Error  & GCG & NGCG & & GCG & NGCG   & &GCG & NGCG \\
 \hline
$\sigma(A_s)$
                   & $0.0220$
                   & $-$&
                   & $0.0510$
                   & $-$
                   && $0.0190$
                   & $-$\\

$\sigma(\beta)$
                    & $0.0750$
                   & $0.0097$&
                   & $0.2350$
                   & $0.3750$
                   && $0.0630$
                   & $0.0087$\\

$\sigma(w)$
                    & $-$
                   & $0.0450$&
                   & $-$
                   & $0.3550$
                   && $-$
                   & $0.0270$\\

$\sigma(H_0)$
                    & $0.4000$
                   & $0.8700$&
                   & $0.6500$
                   & $0.2600$
                   && $0.3400$
                   & $0.2300$\\

$\sigma(\Omega_{de})$
                    & $-$
                   & $0.0078$&
                   & $-$
                   & $0.1600$
                   && $-$
                   & $0.0032$\\
\hline\hline
\end{tabular}
\caption{\label{tab2} Constraint errors for cosmological parameters of the $\Lambda$CDM, $w$CDM, CPL, $\alpha$DE, GCG, and NGCG models by using the CMB+BAO+SN, GW, and CMB+BAO+SN+GW data.}
\centering
\end{table*}

\begin{table*}[htbp]
\small
\setlength\tabcolsep{2.8pt}
\renewcommand{\arraystretch}{1.5}
\centering
\begin{tabular}{cccccccccccc}
\\
\hline\hline &\multicolumn{2}{c}{CMB+BAO+SN}&& \multicolumn{2}{c}{GW}&& \multicolumn{2}{c}{CMB+BAO+SN+GW} \\
 \cline{2-3}\cline{5-6}\cline{8-9}
 Accuracy & $\Lambda$CDM & $w$CDM & & $\Lambda$CDM & $w$CDM   & &$\Lambda$CDM & $w$CDM \\
\hline
$\varepsilon(\Omega_{\rm m})$
                   & $0.0191$
                   & $0.0250$&
                   & $0.0312$
                   & $0.0863$
                   && $0.0089$
                   & $0.0119$\\

$\varepsilon(H_0)$
                    & $0.0065$
                   & $0.0122$&
                   & $0.0034$
                   & $0.0077$
                   && $0.0028$
                   & $0.0053$\\

$\varepsilon(w)$
                    & $-$
                   & $0.0336$&
                   & $-$
                   & $0.1214$
                   && $-$
                   & $0.0227$\\
               \cline{2-3}\cline{5-6}\cline{8-9}
Accuracy & CPL & $\alpha$DE && CPL & $\alpha$DE && CPL & $\alpha$DE \\
\hline
$\varepsilon(\Omega_{\rm m})$
                  & $0.0253$
                   & $0.0244$&
                   & $0.1783$
                   & $0.0896$
                  & & $0.0179$
                   & $0.0138$\\

$\varepsilon(H_0)$
                   & $0.0122$
                   & $0.0121$&
                   & $0.0151$
                   & $0.0090$
                   && $0.0075$
                   & $0.0066$\\

$\varepsilon(\alpha)$
                   & $-$
                   & $1.9286$&
                   & $-$
                   & $7.7857$
                   && $-$
                   & $1.5703$\\

$\varepsilon(w_0)$
                   & $0.0836$
                   & $-$&
                   & $0.2644$
                   & $-$
                   && $0.0704$
                   & $-$\\

$\varepsilon(w_a)$
                   & $3.1500$
                   & $-$&
                   & $1.4160$
                   & $-$
                   && $2.8333$
                   & $-$\\

\cline{2-3}\cline{5-6}\cline{8-9}
Accuracy  & GCG & NGCG & & GCG & NGCG   & &GCG & NGCG \\ \hline

$\varepsilon(A_s)$
                   & $0.0302$
                   & $-$&
                   & $0.0696$
                   & $-$
                   && $0.0261$
                   & $-$\\

$\varepsilon(\beta)$
                    & $8.3333$
                   & $3.3448$&
                   & $4.7000$
                   & $1.7045$
                   && $7.0000$
                   & $3.3462$\\

$\varepsilon(w)$
                    & $-$
                   & $0.0449$&
                   & $-$
                   & $0.2689$
                   && $-$
                   & $0.0269$\\

$\varepsilon(H_0)$
                    & $0.0059$
                   & $0.0128$&
                   & $0.0096$
                   & $0.0038$
                   && $0.0050$
                   & $0.0034$\\

$\varepsilon(\Omega_{de})$
                    & $-$
                   & $0.0113$&
                   & $-$
                   & $0.2759$
                   && $-$
                   & $0.0047$\\
\hline\hline
\end{tabular}
\caption{\label{tab3} Constraint accuracies for cosmological parameters of the $\Lambda$CDM, $w$CDM, CPL, $\alpha$DE, GCG, and NGCG models by using the CMB+BAO+SN, GW, and CMB+BAO+SN+GW data.}
\centering
\end{table*}

In this section, we report and discuss the constraint results of the cosmological parameters by using the GW data, compared to the conventional cosmological EM data. We will show how the GW data, as $standard$ $sirens$, help improve the cosmological parameter estimation.

The one-dimensional marginalized posterior distributions and the two-dimensional contours from GW, CMB+BAO+SN, and CMB+BAO+SN+GW are shown in Figs.~\ref{fig1}--\ref{fig6}. The complete constraint results are shown in Table~\ref{tab1}. In Table~\ref{tab2}, we give the constraint errors for the cosmological parameters. Moreover, the constraint accuracies are shown in Table~\ref{tab3}. Note that the error $\sigma$ here is the average of $\sigma_+$ and $\sigma_-$, and for a parameter $\xi$, its accuracy $\varepsilon(\xi)$ is defined as $\varepsilon(\xi) = \sigma(\xi)/\xi$.

Note also that there are some random factors in the simulation of the GW data, which will lead to the situation that the mock GW data produced in two independent processes would be somewhat different and the cosmological parameter estimations from them are not in exact accordance. Therefore, although the analysis for the $\Lambda$CDM model and the $w$CDM model has been done recently in Ref.~\cite{Zhang:2019ple}, in this work we still redo this analysis for the self-consistent and self-contained purposes.


In Fig.~\ref{fig1}, we show the constraint results of the $\Lambda$CDM model. We have the constraint acuracies: $\varepsilon(\Omega_{\rm{m}})=3.12\%$ and $\varepsilon(H_0)=0.34\%$ from GW, $\varepsilon(\Omega_{\rm{m}})=1.91\%$ and $\varepsilon(H_0)=0.65\%$ from CMB+BAO+SN, and $\varepsilon(\Omega_{\rm{m}})=0.89\%$ and $\varepsilon(H_0)=0.28\%$ from CMB+BAO+SN+GW. We can clearly see that the parameter degeneracy directions of GW and CMB+BAO+SN are rather different, and thus the parameter degeneracy in the conventional CMB+BAO+SN constraint can be broken by including the GW data. We find that the constraint on $\Omega_{\rm{m}}$ is improved by 53.4\%, and the constraint on $H_0$ is improved by 56.9\%, by adding the GW data in the cosmological fit.


In Fig.~\ref{fig2}, we show the constraint results of the $w$CDM model. We can see that, compared to the CMB+BAO+SN data, the GW data alone can constrain $H_0$ much better, but can constrain $\Omega_{\rm{m}}$ and $w$ much worse. However, since the parameter degeneracy directions of GW and CMB+BAO+SN are all rather different, the constraints on all the parameters are significantly improved by including the GW data in the fit. For example, for the parameter $w$, we have $\varepsilon(w)=3.36\%$ from CMB+BAO+SN and $\varepsilon(w)=2.27\%$ from CMB+BAO+SN+GW. Thus, the measurement of $w$ is improved by 32.4\% by considering the GW data in the cosmological fit.


In Fig.~\ref{fig3}, we show the constraint results of the CPL model. We can see that the GW data alone can only provide rather weak constraints on the CPL model. However, due to the parameter degeneracies being broken by the GW data, the parameter constraints are still improved by considering the GW data in the fit. For the parameters $w_0$ and $w_a$, we have the constraint errors: $\sigma(w_0)=0.083$ and $\sigma(w_a)=0.315$ from CMB+BAO+SN, and $\sigma(w_0)=0.070$ and $\sigma(w_a)=0.255$ from CMB+BAO+SN+GW. Therefore, the constraints on $w_0$ and $w_a$ are improved by 15.7\% and 19.0\%, respectively, by including the GW data.



In Fig.~\ref{fig4}, we show the constraint results of the $\alpha$DE model. For the $\alpha$DE model, the limit of $\alpha = 1$ corresponds to the Dvali-Gabadadze-Porrati (DGP) braneworld model \cite{a:2000}, and the limit of $\alpha = 0$ corresponds to the $\Lambda$CDM model. From the figure, we can clearly see that the DGP model has been convincingly excluded by the current observations. The fit result of $\alpha\simeq 0$ indicates that the $\Lambda$CDM limit of this model is strongly favored by the current observations. We can also see that the GW data alone can only provide a rather weak constraint on the parameter $\alpha$. But the combined CMB+BAO+SN+GW data can constrain $\alpha$ tightly. For the constraints on $\alpha$, we have the results: $\sigma(\alpha)=0.135$ from CMB+BAO+SN, $\sigma(\alpha)=0.545$ from GW, and $\sigma(\alpha)=0.101$ from CMB+BAO+SN+GW. We find that the constraint on $\alpha$ is improved by 25.2\% by including the GW data.

In Figs.~\ref{fig5} and \ref{fig6}, we show the constraint results of the GCG and NGCG models, respectively. The GCG model can be viewed as a model of vacuum energy interacting with cold dark matter, with the $\beta=0$ corresponding to the $\Lambda$CDM model and $\beta=1$ corresponding to the CG model. The NGCG model can be viewed as a model of constant $w$ dark energy interacting with cold dark matter, with $\beta=0$ corresponding to the $w$CDM model. The constraint results of $\beta\simeq 0$ and $w\simeq -1$ indicates that the $\Lambda$CDM limit of these models is strongly favored by the current observations. Still, we can see that the GW data can significantly help improve the constraints on the parameters of these models. For example, for the parameters $A_s$, $\beta$ (in GCG) and $w$ (in NGCG), we have the results: $\sigma(A_s)=0.022$, $\sigma(\beta)=0.075$, and $\sigma(w)=0.045$ from CMB+BAO+SN, and $\sigma(A_s)=0.019$, $\sigma(\beta)=0.063$, and $\sigma(w)=0.027$ from CMB+BAO+SN+GW. We find that the constraints on $A_s$, $\beta$, and $w$ can be improved by 13.6\%, 16.0\%, and 40.0\%, respectively, by considering the GW data in the cosmological fit.

From the results obtained in this work, we find that the simulated GW standard siren data from the ET can tremendously improve the constraints on the cosmological parameters for all the considered dark energy models. This conclusion is quite solid because it is based on the analysis for various dark energy models.

\section{Conclusion}\label{sec4}

In this work, we use the simulated GW standard siren data from the future observation of the ET to constrain the various dark energy cosmological models, including the $\Lambda$CDM, $w$CDM, CPL, $\alpha$DE, GCG, and NGCG models. We also use the current mainstream cosmological probes based on the EM observations, i.e., CMB+BAO+SN, to constrain these models.

We find that the GW standard siren data could tremendously improve the constraints on the cosmological parameters for all these dark energy models. In all the cases, the GW data can be used to break the parameter degeneracies generated by the CMB+BAO+SN data. Therefore, it is expected that the future GW standard siren observation from the ET would play a crucial role in the cosmological parameter estimation in the future.



\begin{acknowledgments}

This work was supported by the National Natural Science Foundation of China (Grants Nos.~11975072, 11875102, 11835009, 11690021, and 11522540) and the National Program for Support of Top-Notch Young Professionals.

\end{acknowledgments}

\end{document}